\documentclass[final,5p,twocolumn]{elsarticle}

\usepackage[mediumspace,mediumqspace,Grey,squaren]{SIunits}
\usepackage{natbib}
\usepackage{subfig}
\usepackage{lineno,hyperref}
\modulolinenumbers[5]

\begin{document}

\begin{frontmatter}

\title{Enhancing spatial resolution of soft x-ray CCD detectors \\
by single-photon centroid determination}
\author[1,2]{A. Amorese}
\author[1]{G. Dellea} \author[1]{L. Braicovich} \author[1]{G. Ghiringhelli\corref{mycorrespondingauthor}}
\address[1]{CNR/SPIN and Dipartimento di Fisica, Politecnico di Milano, Piazza Leonardo da Vinci 32, 20133 Milano, Italy}
\address[2]{European Synchrotron Radiation Facility, 71 Avenue des Martyrs, Grenoble F-38043, France}
\cortext[mycorrespondingauthor]{Corresponding author}
\ead{giacomo.ghiringhelli@polimi.it}

\begin{abstract}
In Charge Coupled Device (CCD) detectors the electrons excited upon absorption of a single x-ray photon quickly diffuse and generate charge-spots often larger than pixel dimensions. In the soft x-ray range this phenomenon drastically limits the effective spatial resolution to approximately 25{\micro\meter}, irrespective of the pixel size. For very low fluence the charge-cloud centroid determination can be used, on each individual spot, to estimate the actual photon impact position with sub-pixel precision. The readout noise and speed, together with the charge and spatial undersampling, are the main factors limiting the accuracy of this procedure in commercial devices. We have developed and extensively tested an algorithm for efficient centroid reconstruction on images acquired by low noise detectors not designed for single photon counting. We have thus measured a position uncertainty of 6-7{\micro\meter} in CCDs with 13.5{\micro\meter} and 20.0{\micro\meter} pixel size, around 1 keV photon energy. We have analyzed the centroid calculation performances by modelling the spot generation process. We show how the resolution is affected by both random uncertainty, mainly ascribed to the readout noise, and systematic error, due to the undersampling of the charge spot. This study was motivated by the growing need of of high-resolution high-sensitivity 2D position sensitive detectors in resonant inelastic (soft) x-ray scattering (RIXS), an emergent synchrotron radiation based spectroscopy.
\end{abstract}


\end{frontmatter}


\section{Introduction}

\begin{figure}
\centering
\includegraphics[width=0.95\columnwidth]{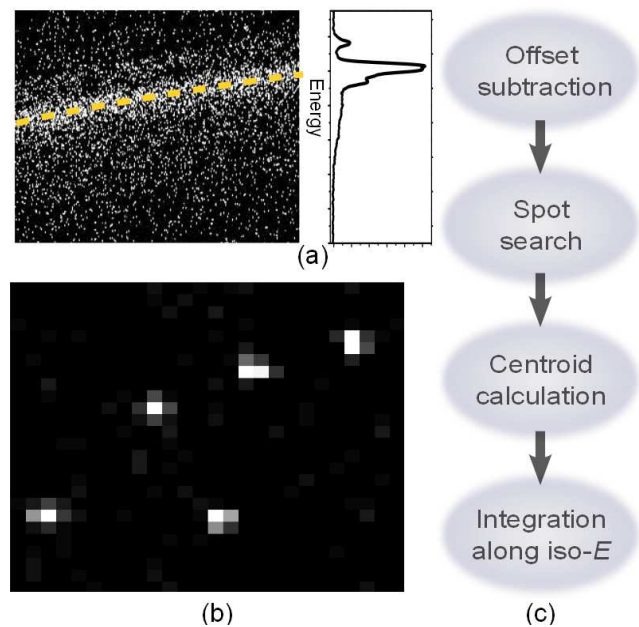}
\caption{(a) Typical image acquired by a CCD detector in a soft x-ray spectrometer (specifically: AXES spectrometer, beamline ID08 of ESRF \cite{dinardo2007gaining}); the integration along the direction indicated by the yellow line leads to the spectrum. (b) Examples of spots produced by individual 930 eV photons on a CCD with 13.5{\micro\meter} pixel size. (c) Simplified flowchart of the algorithm developed for the centroid reconstruction.}
\label{fig:Fig1}
\end{figure}

Third generation synchrotron radiation sources and, in the near future, free-electron lasers are boosting a panoply of x-ray scattering techniques: high resolution diffraction, coherent scattering and holographic imaging, small angle scattering, resonant and non-resonant inelastic scattering. Charge Coupled Device (CCD) cameras are probably the most common 2D position-sensitive detectors for scientific applications and their spatial resolution is a key parameter in several applications, including those based on x rays up to few keV in energy. Usually in elastic scattering (including imaging) the signal is relatively intense and large dynamic range and short read-out time are the most important properties of a detector. Different is the case of inelastic scattering spectroscopy, usually limited by a too low counting rate that calls for an optimization of the sensitivity through the optimization of the quantum efficiency and the minimization of the read-out noise.Moreover the detector ability to capture the impact position is also required. These considerations hold particularly for resonant inelastic x-ray scattering (RIXS) in the soft x-ray range \cite{amnt2011reviewRIXS}\cite{braicovich2010magnonsLSCO}. In modern soft-RIXS spectrometers photons are dispersed \emph{vs} their energy (or wavelength) by a grating ruled on a ultra-high quality grazing incidence mirror, and so that their impact position on the detector surface translates directly into photon energy \cite{dallera1996AXES}\cite{Hague2005flatfieldspectrometer}\cite{dinardo2007gaining}\cite{ghiringhelli2006saxes}\cite{strocov2011grating}\cite{Lai2014taiwanRIXS}. Therefore the image acquired by the CCD camera can be interpreted as a recording of the spectral intensity, organized in parallel isoenergetic lines: the actual spectrum is obtained by integrating along that direction, as depicted in Figure~\ref{fig:Fig1}(a). The detector spatial resolution therefore contributes to the spectrometer energy resolution, together with optical aberrations, grating shape errors and source size. As mentioned above in high-resolution RIXS experiments, when all settings are optimized for minimizing the instrumental bandwidth, the photon flux reaching the detector is extremely low, of the order of 1 ph {\milli\meter}$^{-2}${\second}$^{-1}$ in the most intense part of the spectrum, as low as 3 orders of magnitude lower in the less intense spectral regions and up to 10 times higher in exceptional cases. These numbers explain why the ultimate sensitivity in RIXS is set by the dark-current and the read-out noise of the detector, which is thus usually read very slowly after an exposure time of 1 to 5 minutes. In the presently leading RIXS instruments \cite{dinardo2007gaining}\cite{ghiringhelli2006saxes}\cite{Lai2014taiwanRIXS} the detector is always a back-illuminated CCD, cooled below -55$^{\circ}$C, with a few cm$^2$ active area and pixel lateral size ranging 10{\micro\meter} to 20{\micro\meter}.\\

The resolution of commercial thinned back-illuminated soft x-ray CCDs has been shown to be about 25{\micro\meter}\cite{ghiringhelli2006saxes} and almost independent on the physical pixel size \cite{dinardo2007gaining}. In fact, a photon absorbed in Si generates, through an Auger cascade and inelastic scattering processes taking place in few {\femto\second}, a number of charges proportional to its energy (3.65 e$^-$/eV). Subsequently the electrons spread by diffusion over an area corresponding to several pixels \cite{hopkinson1983charge}, thus generating an extended spot in the image (Figure~\ref{fig:Fig1}(b)). This limitation is absent in the optical range, where every photon generates at most one low energy electron that doe not diffuse randomly before being trapped inside a pixel, whose size directly determines the CCD spatial resolution. The loss of resolution can be overcome with the use of centroid reconstruction methods, where the photon impact position is estimated from the intensities accumulated in the different pixels of the spot. This method can work only with isolated events, because overlapping spots can be easily misinterpreted. Centroiding has already been used with good result for hard x-rays\cite{abboud2013sub, miyata2003mesh}, whereas only recently it has been explored in the soft x-ray range. The first published tests are very encouraging \cite{soman2011improving, hall2012improving}, but an evaluation of the actual improvement in spectroscopy is still missing. Moreover, recent studies are focusing on the use of innovative detectors specifically designed for single photon counting, which provide a shorter readout time with low noise, at the price of a higher cost \cite{soman2013improving}\cite{Soman2013emCCDforRIXS}.\\

We have investigated the application of a centroid reconstruction algorithm on images acquired by commercial CCD detectors in all similar or identical to those already in use in soft x-ray spectrometers. The actual spatial resolution was measured with photons around 1{\kilo\electronvolt} with a traditional x ray source. Moreover, with the help of simulations, we discuss the critical hardware and software parameters that affect the final resolution, providing a full explanation of their interaction and some useful suggestions for the actual utilization of the centroid reconstruction in RIXS spectrometers.

\section{Methods}

\subsection{Experimental Test}
We tested the effective spatial resolution by projecting the profile of a straight razor blade onto the surface of the CCD and by estimating its width. We used two commercial thinned back-illuminated CCD cameras by Roper Scientific: Princeton PI-SX 1300 (CCD20) and Princeton PI-SX 2048 (CCD13), with an active area made of $1300\times1340$ and $2048\times2048$ squared pixels of 20 and 13.5{\micro\meter} size respectively. In order to reduce the dark current to negligible values, CCD20 was cooled down to $-60\,^{\circ}\mathrm{C}$ with a cascade of Peltier cells, while CCD13 was kept at $-110\,^{\circ}\mathrm{C}$ using liquid nitrogen. The readout noise was minimized to 3-5 electrons rms by using the 100 kHz pixel reading frequency. The centroid calculation requires separated and isolated spots on the image, therefore we acquired images with relatively short exposure times (20\second and 40\second for CCD20 and CCD13 respectively), resulting in a low density of photons in each acquisition. The total number of images acquired is 3240 for CCD20 and 2970 for CCD13, divided in 18 and 33 groups respectively.\\

\begin{figure}
\centering
\includegraphics[width=1\columnwidth]{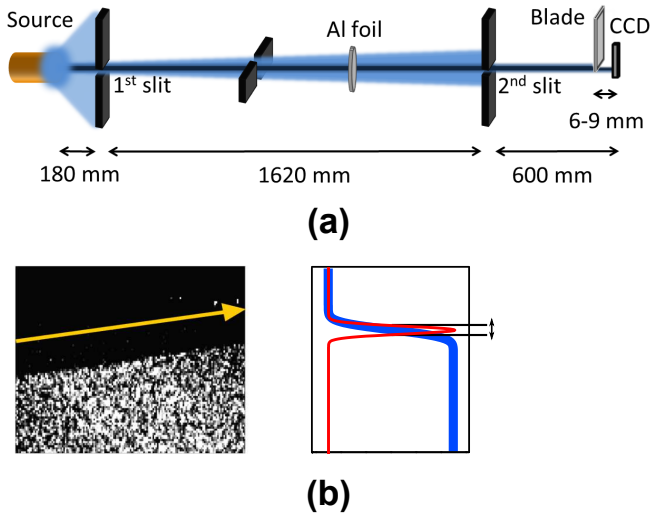}
\caption{(a) Experimental set-up for the resolution tests. A Cu anode produces 930-950 eV photons and two slits collimate the beam, which illuminates a razor blade, imprinting the blade shadow on the images acquired by the CCD. (b) Sum of 360 acquisitions of CCD13. It is visible the shadow of the blade edge, along which the image is integrated to get the the intensity profile, similar to an error function. The FWHM of the underlying Gaussian function gives our resolution estimate.} \label{fig:Fig2}
\end{figure}

The experimental set-up is depicted in Figure~\ref{fig:Fig2}(a). We used a conventional copper anode x-ray source, operated at 3 keV, producing a continuous spectrum between 500eV and 2000eV, dominated by the Cu-$L_{\alpha,\beta}$ lines, around 930 eV. A collimated beam is obtained with two adjustable slits (40{\micro\meter} opening). Part of the beam is shadowed by a razor blade, located few millimetres before the CCD surface, producing an image as shown in  Figure~\ref{fig:Fig2}(b) (sum of 360 images acquired by CCD13). The images are then integrated along channels parallel to the blade edge, resulting in an intensity profile that can be fit with an error-function (Figure~\ref{fig:Fig2}(b)). The derivative of the fitting function is a Gaussian, whose full width at half maximum (FWHM) is the estimated resolution.
The blurring of the images due to the diffraction from a knife edge was minimized by setting the distance blade-camera to 6mm and 9mm for CCD20 and CCD13 respectively, with an estimated effect on the resolution of about 2{\micro\meter}. The broadening due to the residual beam divergence is less than 0.2{\micro\meter}.

\subsection{Center of mass algorithm}
The algorithm searches the spots on the image and determines the positions of the centroids by a center of mass calculation. Its scheme is depicted in Figure~\ref{fig:Fig1}(c). After a constant offset subtraction, the possible central pixels of the spots are selected according to their intensity being included in a range optimized empirically. An $N \times N$ matrix ($N=3,5$) is selected around each of those ``central" pixels of the spots and duplicated spots are discarded. For each spot the centroid coordinates are calculated as first order momentum.

The positions of the spot centroids are thus known with sub-pixel accuracy. Then the integration along a direction usually not parallel to the pixel rows can be performed to obtain the intensity profile (a spectrum as in Figure~\ref{fig:Fig1}(a) or a blade edge as in Figure~\ref{fig:Fig2}(b)). This integration is made by sampling with sub-pixel channels whose width is a free parameter.
We estimated the actual resolution enhancement for spectroscopic applications by comparing the intensity profile resulting from our algorithm with the one obtained with an algorithm that does not perform the centroid reconstruction. This algorithm is the one currently in use to process the images acquired with the SAXES spectrometer \cite{ghiringhelli2006saxes}, mounted on the ADRESS beam line of Swiss Light Source, and the AXES spectrometer, on the beam line ID08 of the European Synchrotron Radiation Facility. The details about this algorithm can be found in Ref\cite{dinardo2007gaining}.

\subsection{Simulations}

The simulations provide a comparison with the experimental data and a tool to study how the centroid calculation is affected by hardware properties and software parameters.  We generated virtual spots by simulating the splitting of the charge clouds among adjacent pixels. The average shape of a charge cloud on the detector surface was modeled with a 2D Gaussian function of 25{\micro\meter} FWHM.
The Gaussian shape model was calculated by Hopkinson \cite{hopkinson1983charge, hopkinson1987analytic} and later confirmed by experiments \cite{hirago1998big} \cite{hall2012improving}. Pavlov and Nousek\cite{pavlov1999charge} predicted some shape modifications if the photon is absorbed deep in the field free region of the substrate, but in the thinned detectors under test, where this region is almost completely etched out \cite{princeton} we do not expect these deviations to be present. Therefore the Gaussian model was considered suitable. The width of the Gaussian function, instead, can vary depending on some detector characteristics, one of these being the internal fields. Our model uses the value measured by Dinardo \cite{dinardo2007gaining} and Ghiringhelli \cite{ghiringhelli2006saxes} on Princeton detectors of the same family of those tested in our experiment. Each spot is constituted by a number of electrons that is randomly chosen using a Gaussian distribution, centered in $N_{el}=E (eV)/3.65$ and with a standard deviation $\sigma=\sqrt{0.115 \times N_{el}}$, being 0.115 the Fano factor for Si. The electrons are randomly assigned to the pixels around the photon impact position according to the 2D Gaussian shape of the cloud.\\ 

This procedure was used for two types of analysis. We produced virtual images similar to the experimental ones, with monochromatic photons (930 eV) randomly distributed in a region delimited by an inclined line, which represents the blade edge. These images were analysed in the same way as the experimental images. Moreover, we performed an analysis of the possible errors in the centroid reconstruction. We calculated the centres of mass of 3000 virtual spots centred in the same assigned position inside a pixel, and this for a fine grid of positions to cover 1 quadrant of the pixel. The systematic error was determined by analysing the difference between the mean position of the centroids and the actual centres of the spots, while the random error was calculated as the standard deviation of the centroids positions. We repeated these calculations centring the 3000 spots in the points of a regular grid inside a quarter of a pixel, obtaining a map of the errors depending on the impact position of the photon (after copying the results to the other quadrants to cover the whole pixel).

\section{Results}

\subsection{Resolution enhancement}

\begin{figure}
\centering
\includegraphics[width=0.85\columnwidth]{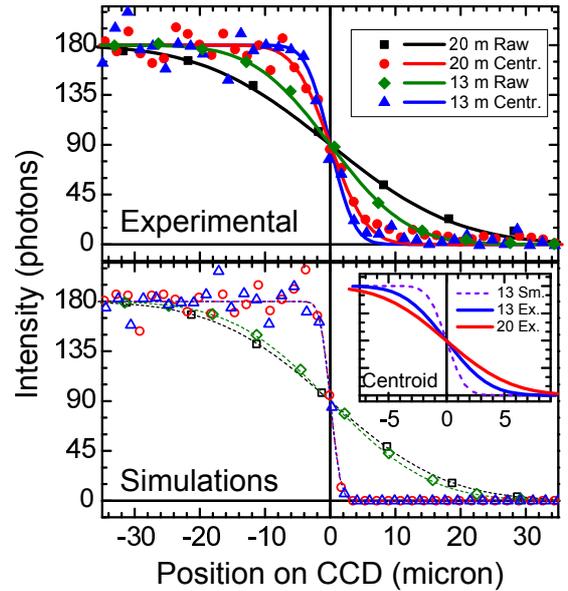}
\caption{Intensity profiles from experimental (top panel) and simulated (bottom panel) images, without (``Raw") and with the centroid reconstruction (``Centroid"). Inset: Comparison of the resolution achieved with experiments and simulations. The fits are represented by continuous lines. The intensity scales refer to the profiles elaborated with the centroid reconstruction algorithm, slightly rescaled (less than $5\%$) to be superimposed . The resolution enhancement is about $3.5\times$ in the experimental case, achieving 9.8{\micro\meter} and 6.8{\micro\meter} respectively for CCD20 and CCD13, while the simulations (without noise) show a resolution about 2.8-2.9{\micro\meter} for both pixel sizes.} \label{fig:Fig3}
\end{figure}

The experimental tests (Figure~\ref{fig:Fig3} top panel) proved that an actual resolution enhancement of about 3.5 times is achievable thanks to the centroid reconstruction algorithm, while the analysis of simulated images (bottom panel) showed that without the experimental uncertainties the enhancement is even higher.\\

By processing the experimental images without centroid reconstruction, we get a resolution estimate of 36{\micro\meter} and 22{\micro\meter} for CCD20 and CCD13 respectively. Both these values differ from the cloud size of 25{\micro\meter}, meaning that other contributions must be taken into account; one of these being the pixel size, that directly influences the sampling of the final profile.
When the centroid reconstruction algorithm is used, the resolutions estimated are 9.8{\micro\meter} and 6.8{\micro\meter} for the images acquired by CCD20 and CCD13. These results were obtained using matrices of $3\times3$ and $5\times 5$ pixels for the centroid calculation and channel widths of 1/11 and 1/7 of pixel respectively for the two detectors.\\

To have a direct comparison with the experimental data, we produced simulated images with pixel sizes of 20{\micro\meter} and 13.5{\micro\meter}. The resolution estimated without centroid reconstruction is 33{\micro\meter} and 28{\micro\meter} respectively for the two cases, labelled in Figure 3, bottom panel, as “CCD20” and “CCD13”. These values confirm the difference between the actual resolution and the expected 25{\micro\meter} of the cloud size even in the ideal case of simulations, where the noise and other experimental uncertainties are not present. The profiles obtained after the centroid elaboration are almost perfectly superimposed, resulting in a resolution of about 2.8{\micro\meter} for the 20{\micro\meter} pixel images and 2.9{\micro\meter} for the 13.5{\micro\meter} pixel. These values further confirm the resolution enhancement achievable by the centroid reconstruction algorithm. It must be noted that these values are comparable with the width of the channels used for the image integration, 1.82{\micro\meter} and 1.91{\micro\meter} respectively. The analysis performed on simulated images using tighter integration channels (1/20 of the pixel) proved that a resolution of 1.8{\micro\meter} can be achieved for the 13.5{\micro\meter} detector. However, such a fine discretization of the profile would unlikely be used in a real experiment because it implies an drastic loss of counts per data-point, due to the division of the same amount of photons among a higher number of channels.

\subsection{Intensity modulation after elaboration}

The images processed by the centroid reconstruction algorithm show, in certain cases, a regular modulation of the number of photons, which is not present on the raw acquisitions. This modulation is related to a non-uniform distribution of the centroids within the pixel after the elaboration.\\

\begin{figure}
\centering
\includegraphics[width=0.9\columnwidth]{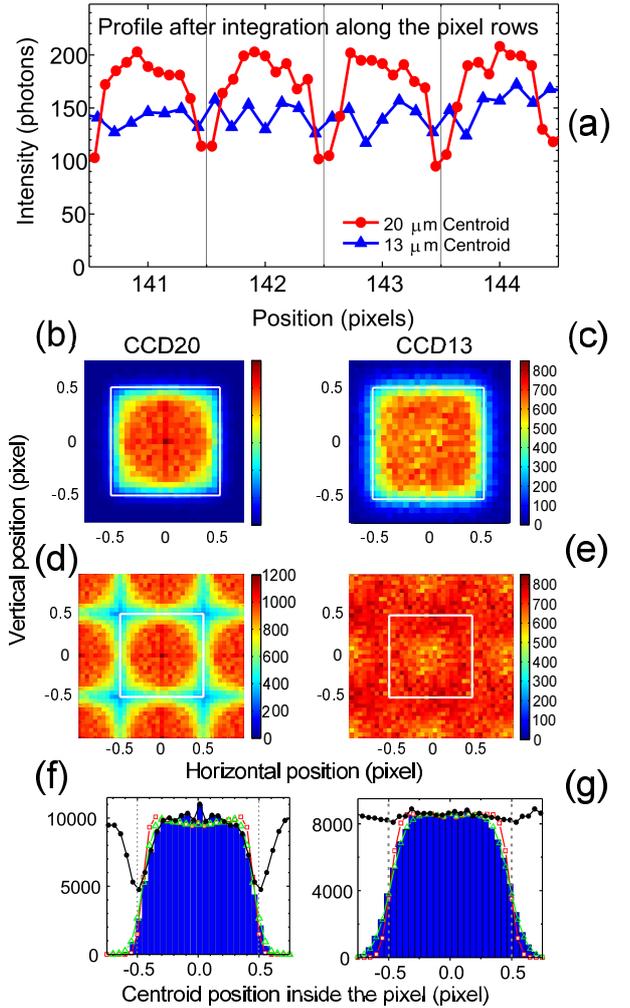}
\caption{Statistical study of the centroid calculated positions for the two detector used in this work. For the CCD20 (CCD13) a $3 \times 3$ ($5 \times 3$) matrix was used for centroiding. (a) Experimental intensity profile of a uniformly illuminated region of the image obtained by integration along horizontal lines after centroid reconstruction. (b and c) Experimental average distribution of the calculated centroid positions inside a single pixel, when reassigning all events of the image to the same central pixel.. (d and e) Average intensity patterns after the centroid reconstruction obtained by reproducing theintensity pattern of panels b and c in all pixels. (f and g) Intensity profiles obtained by integration from panels b-e. Blue histograms are the profiles obtained from panels b and c, black circles those of panels d and e. The corresponding simulated distributions are shown with red squares and green triangles, respectively without and with noise (5 electrons rms for CCD20 and 3 electrons rms for CCD13). The intensity modulations with physical pixel periodicity are thus present only in the CCD20 case (panels a and f), because the $3 \times 3$ matrix is too small to cover the whole charge distribution, whose tails are discarded leading to a systematic error in the centroid determination.}\label{fig:Fig4}
\end{figure}

The intensity profiles shown in Figure~\ref{fig:Fig4}(a) are obtained by integrating a uniformly illuminated region of the images along horizontal channels parallel to the rows of pixels. The profile calculated using CCD20 images is characterized by deep minima on the edges of the pixels, while the other line (obtained from CCD13 images) does not show this modulation.\\

To study the the center of mass average distribution inside the pixels we arranged all the centroids in a single pixel, according to their positions inside the central pixel of the matrix used for the centroid calculation.
The resulting intensity maps (depicted in Figure~\ref{fig:Fig4}(b, c) for the two detectors) give the average pixel response to uniform illumination (random photons impact positions inside the pixels). By reproducing these distributions in the neighbouring pixels, we obtained the average pattern resulting from the elaboration of uniform images  (Figure~\ref{fig:Fig4}(d,e)). In these maps, the total intensity around the boundary lines between two pixels is given by the sum of the centroids coming from the two confining pixels. Finally, by integrating the intensity maps of Figure~\ref{fig:Fig4}(b) along one direction we obtained the distribution of the centroids shown in the histograms of Figure~\ref{fig:Fig4}(f, g) (blue hisograms). By repeating these shapes in the adjacent pixels we produced the average intensity profiles, depicted with the black lines with circles. \\

In an ideal case, with uniform illumination, we would expect the distribution to be uniform inside the pixel and to drop to zero at the edges. Because of the statistical noise, in the distributions of Figure~\ref{fig:Fig4}(c,e,g) some centres near the edges are assigned out of the pixel and the intensity drop is gradual. The decreasing is symmetrical across the borders of the pixel in the CCD13 distribution (right) and therefore, after the sum with the intensity coming from the adjacent pixel, the resulting image (Figure~\ref{fig:Fig4}(c)) and profile (Figure~\ref{fig:Fig4}(g), black circles) are uniform.
On the other hand, the distribution of the centroids in the pixel of CCD20 (Figure~\ref{fig:Fig4}(b,d,f) is narrower and decreases more steeply inside the pixel, with a low intensity on the edges. Therefore, the average image and the intensity profile (Figure~\ref{fig:Fig4}(b) and black circles in Panel (g)) are not uniform, and show a depletion of the events on the boundary lines between pixels.\\	

These centroid distributions are confirmed by the results of the simulations. In Figure\ref{fig:Fig4}(f,g), the simulations with noise (green empty triangles) agree perfectly with the experimental data, confirming the soundness of the model used for the spot generation algorithm. The noise added to pixels is 5 electrons rms for the simulations with 20{\micro\meter} pixels (left) and 3 electrons rms for those with 13.5{\micro\meter} pixels (right).
Without noise, both the centroid distributions (red empty squares) show a larger plateau in the center and a sharper drop at the boundaries. However, the intensity is still symmetrical across the pixel edges for the 13.5{\micro\meter} pixel (right), ensuring a uniform intensity in the final profile, while only few centroids are present near the edges of the 20{\micro\meter} pixel (left), thus confirming the problem of the intensity modulations in this case.
Therefore, a systematic error is always present with 20{\micro\meter} pixels, while this problem does not affect the centroid reconstruction on 13.5{\micro\meter} pixels. Apart from the pixel size, the main difference between the two cases is the size of the matrix of pixels used for the centroid calculation (respectively, $3\times3$ and $5\times 5$ pixels). To further investigate the causes of the systematic error, we used the simulations to study the centroid calculation performances in function of the size of the matrix, the size of the pixels and the amount of noise inside pixels. All the results are described and discussed in the next section.

\subsection{Analysis of the systematic and random errors of centroid reconstructions}

\paragraph{Systematic and random errors}

The parameter that mostly influences the distribution of the centroids in the pixel, and therefore the intensity modulations, is the size of the matrix used for the center of mass calculation. Too small matrices leave out part of the charge cloud, causing a systematic error that shifts the centroids towards the pixel center.\\

\begin{figure}
\centering
\includegraphics[width=0.95\columnwidth]{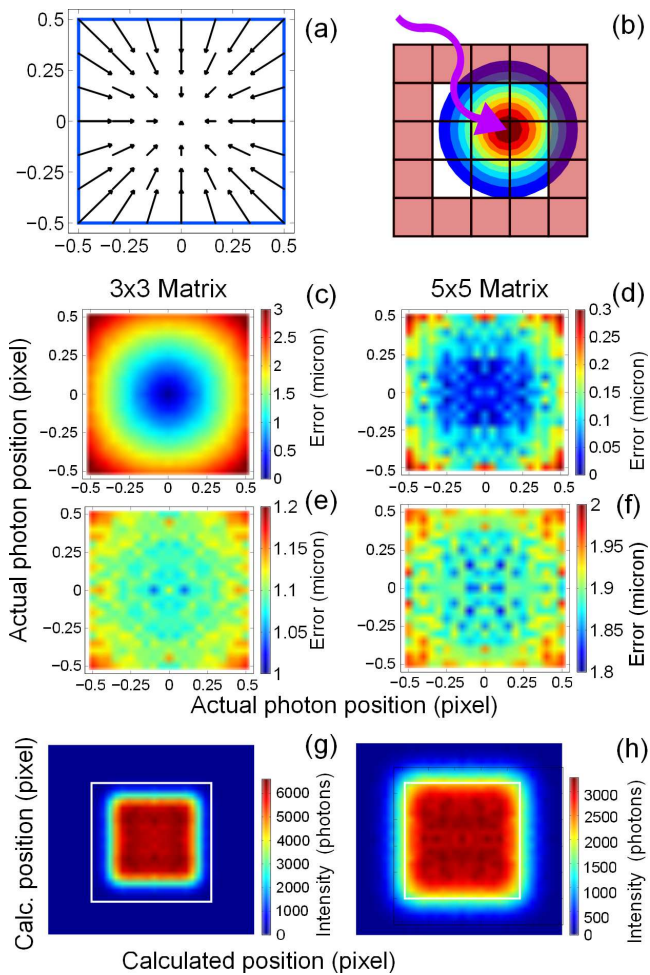}
\caption{Simulated systematic error made when excluding a portion of the charge cloud from the centroid calculation. Data refer to simulations with 13,5{\micro\meter} pixels and 3 electrons rms of artificial noise. (a) Shift of the calculated centroid positions from the actual spot centre position, using a $3\times3$ matrix. (b) Representation of a Gaussian cloud of 25{\micro\meter} on 13.5{\micro\meter} pixels. A portion of the cloud falls out of the $3\times3$ matrix of pixels and is collected by the red-shaded pixels, which are instead considered using a $5\times 5$ matrix. (c,d) and (e,f) Systematic and random error as function of the actual spot centre position inside the pixel. (g,h) Corresponding centroid distribution in the pixels. The $5\times 5$ matrix (panels d,f,h; right column) guarantees a zero systematic error and a uniform distribution of the centroids in the pixels, at the cost of a larger random error due to the increased readout noise associated to a bigger number of pixels (25) with respect to with the $3\times 3$ matrix (left column).} \label{fig:Fig5}
\end{figure}

In Figure~\ref{fig:Fig5}(b) a Gaussian cloud of 25{\micro\meter} FWHM is drawn on a matrix of 13.5{\micro\meter} pixels. The wider matrix of $5\times 5$ pixels includes practically the whole charge cloud, while, neglecting the outer pixels (coloured in red) and using only a $3\times 3$ matrix, part of the cloud is not considered. This affects the centre of mass calculation especially when the photon is absorbed close to the central pixel edge, because a significant portion of the cloud is excluded from the matrix only on one side. The error made using a $3\times 3$ matrix is shown in Figure~\ref{fig:Fig5}(a), which depicts the mean displacement of the calculated centroids from the actual spot centre position. The systematic shift points always towards the centre of the pixel and it grows when approaching the edges.\\

The comparison between the centroid reconstruction performed using a $3\times 3$ and a $5\times 5$ matrix is shown in Figure~\ref{fig:Fig5}(c)-(h), for virtual spots on 13.5{\micro\meter} pixel and with 3 electrons rms of noise. Figure~\ref{fig:Fig5}(c,d) represents the modulus of the systematic error across the pixel, for the $3\times 3$ (left panel) and $5\times 5$ matrix (right panel) case. The maximum systematic error on the corner of the pixel is 3{\micro\meter} using the $3\times 3$ matrix, while it is about 10 times smaller by performing the centroid calculation on a $5\times 5$ matrix of pixels. Regarding the random error (depicted in the maps of Figure~\ref{fig:Fig5}(d)) it is approximately uniform across the pixel, with a slight increase near to the corners. The random error is bigger using a $5\times 5$ matrix (right panel), because it is partly caused by noise inside pixels and, by considering more pixels, the total amount of noise increases.\\
The resulting centroid distributions in the pixel are shown in Figure~\ref{fig:Fig5}(e). The big systematic error made with a $3\times 3$ matrix completely empties the edges of the pixel, shifting the centroids inwards and doubling the intensity in the centre (left panel). On the other hand, the intensity map obtained using a $5\times 5$ matrix (right panel) is uniform inside the pixel and decreases more softly near the edges, due to both the small systematic error and the bigger random error. This distribution is similar to the experimental one of Figure~\ref{fig:Fig4}(b) and does not cause intensity modulations.
Therefore, to prevent from intensity modulations, it should be used an area for the centroid reconstruction that is wide enough to include the whole charge cloud.

\begin{figure}
  \centering
  \includegraphics[width=0.8\columnwidth]{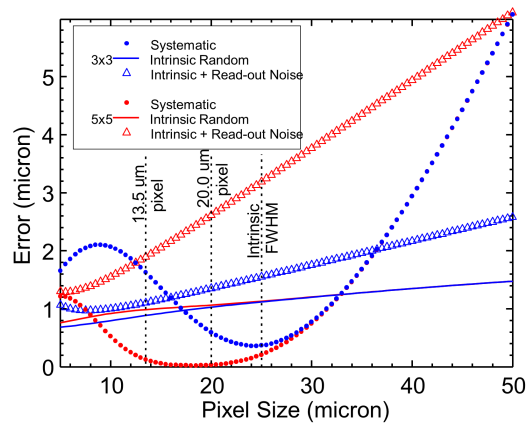}
  \caption{Simulated average systematic and random error in function of the pixel size, with and without 3 electrons rms of artificial noise. For pixels larger than 25{\micro\meter}, the systematic error is caused by a wrong sampling of the spot, while for smaller pixels it is caused by the exclusion of the outer portion of the cloud from the matrix of pixels. The $5\times 5$ matrix gives a zero systematic error for pixel sizes between 12 and 25{\micro\meter}, but due to the large random error caused by noise, it does not always lead to the best resolution.}\label{fig:Fig6}
  \end{figure}

\paragraph{Pixel and matrix size}
Besides the number of pixels in the matrix, the parameter that influences the physical size of the area used for the centroid calculation is the pixel size. To study its effect we generated virtual spots varying the pixel size from 5 to 50{\micro\meter}, in steps of 0.5{\micro\meter}, with and without 3 electrons rms of noise. Figure~\ref{fig:Fig6} shows, as function of the pixel size, the average values of the systematic and random errors within the pixel, using a $3\times 3$ and a $5\times 5$ matrix for the centroid calculation. The black vertical line at 25{\micro\meter} indicates the FWHM of the charge cloud. For pixels larger than this value the systematic error is caused by a wrong sampling of the charge cloud by the pixels and it increases with the pixel size. The results are independent on the size of the matrix of pixels used for the calculation, because the pixels are large enough to include the whole charge cloud with a $3\times3$ matrix. Smaller pixels, on the contrary, imply a smaller area for the centroid calculation, leading to the exclusion of a part of the cloud. Therefore the systematic error for the $3\times3$ matrix grows when reducing the pixel size, while the $5\times 5$ matrix is wide enough to include the whole charge cloud for pixel sizes down to 13{\micro\meter}, with a zero systematic error. Finally, for very small pixels, the systematic error, in general smaller than the pixel size, goes to zero independent on the matrix size used.\\

The random error, instead, it is monotonically increasing with the pixel size. Without noise the random error is around 0.6-1{\micro\meter}, independent on the matrix size, and it varies slowly. This portion of the random error is due to the stochastic nature of the charge splitting among pixels and cannot be cancelled. The addition of the artificial noise strongly affects the centroid calculation on a $5\times 5$ matrix, with an increase of the random error up to 3.1{\micro\meter} for 25{\micro\meter} pixels, while it has a small and almost negligible effect on the $3\times3$ matrix. Therefore, the wider matrix, which in general minimizes the systematic error and prevents from the intensity modulations, implies an increase of the random error that can lead to a worse overall resolution. An example of this trend is given by the two pixel sizes used in our experimental test: passing from 13.5{\micro\meter} to 20{\micro\meter} pixels, the systematic error made using the $3\times3$ matrix decreases and the random error introduced with the $5\times 5$ matrix increases. Therefore, to minimize the overall error, we used a $5\times 5$ matrix for the centroid calculation on 13.5{\micro\meter} pixels and a $3\times3$ matrix on 20{\micro\meter} pixels.

\begin{figure}
\centering
\includegraphics[width=0.95\columnwidth]{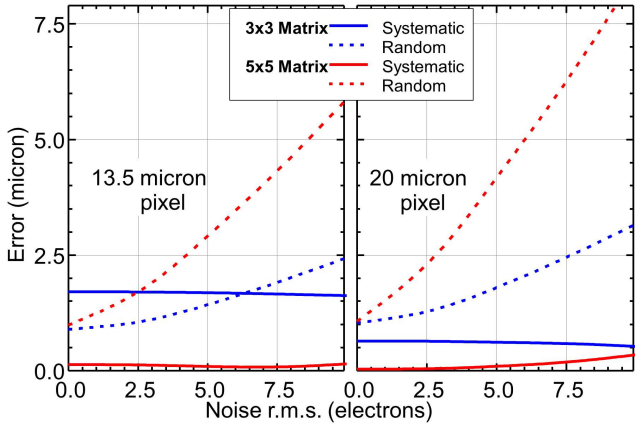}
\caption{Simulated systematic and random error as function of the detector readout noise, for 13.5{\micro\meter} pixels (left) and 20{\micro\meter} (right) and for both the considered matrix sizes. The random error increases rapidly with the noise for the $5\times 5$ matrix in both the cases. The  $3\times 3$ matrix is always preferable with the 20{\micro\meter} pixels, because of the smaller systematic error. With  13.5{\micro\meter} pixels instead, the systematic error for a $3\times 3$ matrix is comparable with the random error and for very low noise detectors a $5\times 5$ matrix is be the best solution.}\label{fig:Fig7}
\end{figure}

 \paragraph{Readout noise}
As discussed above, the readout noise contribution to the random error affects the choice of the best matrix size, thus we investigated the evolution of the errors versus noise for the two matrices and the two pixel sizes considered.
We ran the simulations by varying the noise inside pixels from 0 to 10 electrons rms. Results are shown in Figure~\ref{fig:Fig7} for 13.5{\micro\meter} pixels (left panel) and 20{\micro\meter} pixels (right panel).
In the 20{\micro\meter} pixel case, the resolution is dominated by the random error contribution. In order to minimize the error and to maximize the resolution, a $3\times 3$ matrix is always preferable, being less affected by noise than the $5\times 5$ matrix.
The optimal matrix size for 13.5{\micro\meter} pixel, instead, depends on the amount of noise and it is not trivial to find. Using a $3\times 3$ matrix, the systematic error made is comparable to the random error and their combination can be higher than the overall error made using a $5\times 5$ matrix. In general, a $5\times 5$ matrix optimizes the resolution for very low noise detectors, while for detectors with higher noise, the $3\times 3$ matrix guarantees the smallest overall errors.

\section{Conclusions}

We have shown how centroiding techniques can improve the spatial resolution of soft X-ray CCD detectors. The experimental tests show that the resolution achievable is 6.7{\micro\meter} or even less.
Important suggestions for the future development of detectors and for the application of centroid techniques arise from our simulations. While the current pixel size of 12 - 25{\micro\meter} is suitable to perform a good centroid calculation, the limiting parameter turned out to be the detector readout noise. As a matter of fact, in order to cancel the systematic error and prevent artefacts, a $5\times 5$ matrix of pixels should be used for centroiding instead of a $3\times 3$, but this implies higher noise, with a consequent loss in resolution due to the random error. To overcome this problem, the forthcoming research will address to the study of correction methods to cancel the systematic error, in order to allow the use of smaller matrices and reduce the noise. However, the resolution achieved by our test already provides the needed improvement for the performances of soft x-rays dispersive spectrometers and centroid techniques can be applied using the detectors already available at most facilities.\\

The most important limitation comes from the need of isolated spots, which requires, in the modern soft -ray emission spectrometers, an exposure time of few seconds.
By comparing this time with the detector readout time, of 15-40 seconds for the CCD cameras tested (depending on the actual numer of pixels read), we obtain a very low duty cycle and a consequent low photon detection efficiency. Therefore future tests will investigate the use of faster reading modes which, at the cost of a slightly higher noise, guarantee higher duty cycles and efficiency.\\

\section{Acknowledgments}
The authors would like to thank the staff of the beam line ID08/ID32 of ESRF for providing full support and lending one detector. Insightful discussions with N.B. Brookes are gratefully acknowledged. AA is grateful to Kurt Kummer for the precious suggestions and the support during the drafting of this paper. Part of this work was supported by the PIK project ``POLARIX" of Italian Ministry of Research (MIUR).

\section*{References}

\bibliography{PaperBib}

\end{document}